\documentstyle[epsfig]{mn}

\newif\ifAMStwofonts

\ifoldfss
  \ifCUPmtlplainloaded \else
    \NewTextAlphabet{textbfit} {cmbxti10} {}
    \NewTextAlphabet{textbfss} {cmssbx10} {}
    \NewMathAlphabet{mathbfit} {cmbxti10} {} 
    \NewMathAlphabet{mathbfss} {cmssbx10} {} 
  \fi
  \ifAMStwofonts
    \ifCUPmtlplainloaded \else
      \NewSymbolFont{upmath} {eurm10}
      \NewSymbolFont{AMSa} {msam10}
      \NewMathSymbol{\upi}     {0}{upmath}{19}
      \NewMathSymbol{\umu}     {0}{upmath}{16}
      \NewMathSymbol{\upartial}{0}{upmath}{40}
      \NewMathSymbol{\leqslant}{3}{AMSa}{36}
      \NewMathSymbol{\geqslant}{3}{AMSa}{3E}

      \let\leq=\leqslant 
      \let\geq=\geqslant \let\ge=\geqslant
    \fi
  \fi
\fi 

\ifnfssone
  \newmathalphabet{\mathit}
  \addtoversion{normal}{\mathit}{cmr}{m}{it}
  \addtoversion{bold}{\mathit}{cmr}{bx}{it}
  \newmathalphabet{\mathbfit} 
  \addtoversion{normal}{\mathbfit}{cmr}{bx}{it}
  \addtoversion{bold}{\mathbfit}{cmr}{bx}{it}
  \newmathalphabet{\mathbfss} 
  \addtoversion{normal}{\mathbfss}{cmss}{bx}{n}
  \addtoversion{bold}{\mathbfss}{cmss}{bx}{n}
  \ifAMStwofonts
    \ifCUPmtlplainloaded \else
      \UseAMStwoboldmath
      \makeatletter
      \new@mathgroup\upmath@group
      \define@mathgroup\mv@normal\upmath@group{eur}{m}{n}
      \define@mathgroup\mv@bold\upmath@group{eur}{b}{n}
      \edef\UPM{\hexnumber\upmath@group}
      \new@mathgroup\amsa@group
      \define@mathgroup\mv@normal\amsa@group{msa}{m}{n}
      \define@mathgroup\mv@bold\amsa@group{msa}{m}{n}
      \edef\AMSa{\hexnumber\amsa@group}
      \makeatother
      \mathchardef\upi="0\UPM19
      \mathchardef\umu="0\UPM16
      \mathchardef\upartial="0\UPM40
      \mathchardef\leqslant="3\AMSa36
      \mathchardef\geqslant="3\AMSa3E

      \let\leq=\leqslant 
      \let\geq=\geqslant \let\ge=\geqslant
    \fi
  \fi
\fi 

\ifnfsstwo
  \DeclareMathAlphabet{\mathbfit}{OT1}{cmr}{bx}{it}
  \SetMathAlphabet\mathbfit{bold}{OT1}{cmr}{bx}{it}
  \DeclareMathAlphabet{\mathbfss}{OT1}{cmss}{bx}{n}
  \SetMathAlphabet\mathbfss{bold}{OT1}{cmss}{bx}{n}
  \ifAMStwofonts
    \ifCUPmtlplainloaded \else
      \DeclareSymbolFont{UPM}{U}{eur}{m}{n}
      \SetSymbolFont{UPM}{bold}{U}{eur}{b}{n}
      \DeclareSymbolFont{AMSa}{U}{msa}{m}{n}
      \DeclareMathSymbol{\upi}{0}{UPM}{"19}
      \DeclareMathSymbol{\umu}{0}{UPM}{"16}
      \DeclareMathSymbol{\upartial}{0}{UPM}{"40}
      \DeclareMathSymbol{\leqslant}{3}{AMSa}{"36}
      \DeclareMathSymbol{\geqslant}{3}{AMSa}{"3E}

      \let\leq=\leqslant 
      \let\geq=\geqslant \let\ge=\geqslant
    \fi
  \fi
\fi 

\ifCUPmtlplainloaded \else
  \ifAMStwofonts \else 
    \def\upi{\pi}
    \def\umu{\mu}
    \def\upartial{\partial}
  \fi
\fi

\title[Fornax Spectroscopic Survey]
{The Fornax Cluster Spectroscopic Survey: A Sample
of Confirmed Cluster Dwarfs}
\author[J.H. Deady et al.]
{J.H. Deady$^1$, 
P.J. Boyce$^1$,
S. Phillipps$^1$,
M.J. Drinkwater$^2$, 
A. Karick$^2$,\cr
J.B. Jones$^3$,
M.D. Gregg$^{4,5}$ and 
 R.M. Smith$^6$\\
$^1$Astrophysics Group, Department of Physics, University of Bristol, Bristol BS8 1TL\\
$^2$School of Physics, University of Melbourne, Victoria 3010, Australia\\
$^3$School of Physics and Astronomy, University of Nottingham, University Park,
Nottingham NG7 2RD\\
$^4$Department of Physics, University of California, Davis, CA 95616, USA\\
$^5$ Institute for Geophysics and Planetary Physics, Lawrence Livermore 
National Laboratory, L-413 Livermore, CA 94550, USA\\ 
$^6$Department of Physics and Astronomy, University of Wales Cardiff, 
PO Box 913, Cardiff CF2 3YB}
\date{DRAFT version 0.1, 2002 February 11}

\pagerange{\pageref{firstpage}--\pageref{lastpage}}
\pubyear{2002}

\begin{document}

\maketitle

\label{firstpage}

\begin{abstract}
The  Fornax Cluster Spectroscopic Survey (FCSS) project utilises the Two-degree
field (2dF) multi-object 
spectrograph on the Anglo-Australian Telescope. Its aim is to obtain spectra 
for a complete sample of all 14000 objects with $16.5\leq b_j \leq19.7$ 
{\em irrespective of their morphology} 
in a 12~deg$^{2}$ area centred on the Fornax Cluster. 
 A  sample of 24 Fornax Cluster members has been identified from 
 the first 2dF field (3.1~deg$^{2}$ in area) to be completed. 
  This is the first complete sample of cluster objects of known distance 
  with well defined selection limits.  
  19 of the galaxies (with --15.8$<M_{B}<$--12.7) 
   appear to be conventional dwarf elliptical (dE) 
 or dwarf S0 (dS0) galaxies. The other 5 objects 
 (with  --13.6$<M_{B}<$--11.3) are those galaxies  
   which we described  in Drinkwater  et al. (2000b) and labelled
  `Ultra-Compact Dwarfs' (UCDs). A major result 
 is that the conventional dwarfs  all 
  have   scale-sizes $\alpha$$\ga$3~arcsec ($\simeq$300~pc). 
   This apparent minimum scale size implies an equivalent 
  minimum luminosity  for a  dwarf of a given surface 
 brightness. This produces a  limit on their distribution 
 in the  magnitude-surface brightness plane, such that 
 we do not observe dEs with high surface-brightnesses but  faint
 absolute magnitudes.  Above this  
 observed minimum scale-size  of 3~arcsec, the dEs and dS0s fill the whole area 
 of the magnitude--surface brightness plane sampled by our 
 selection limits.  The observed correlation between 
 magnitude and surface brightness noted by several recent studies 
 of brighter galaxies is not seen with our  fainter cluster sample. 
 A comparison of our results with the Fornax Cluster Catalog (FCC) of Ferguson
 illustrates that attempts to determine cluster membership solely 
 on the basis of observed morphology can produce significant errors. 
   The FCC  identified 17 of the 
   24 FCSS sample (i.e 71 per cent) as being `cluster' members, 
    in particular  missing all 5 of the UCDs. 
   The FCC also suffers from significant contamination:  within the FCSS's field  
    and selection limits, 23 per cent of those objects 
   described as  cluster members by the FCC 
   are shown by the FCSS to be background objects.

\end{abstract}


\section{Introduction}

It is generally perceived that dwarf galaxies are primarily of low 
 surface brightness, and that  lower luminosity
galaxies have fainter surface brightnesses. In particular, low surface brightness
  galaxies (LSBGs) seen towards a cluster are conventionally assumed to
be members, while apparently faint, but high surface brightness
galaxies (HSBGs) are presumed to be luminous objects in the background
(e.g. Sandage, Binggeli \& Tammann 1985).
Indeed, it has frequently been asserted that dwarf galaxies
(in particular) exhibit a strong surface brightness -- luminosity
relation (e.g. Sandage, Binggeli \& Tammann 1985; Bingelli, Sandage \& 
Tammann 1985; Kormendy 1985; Ferguson \& Sandage 1988). 
Binggeli (1994) has particularly persuasively
argued this point with reference to the Local Group. Nevertheless, other
authors have noted that a good deal of the evidence for the correlation
relies on the `eyeball' selection of dwarf galaxies `similar' to ones
already known, that any correlation breaks down at the low luminosity end,
and/or that there may be strong selection effects limiting the detected
galaxies' photometric parameters (e.g. Phillipps, Davies \& Disney 1988;
Irwin et al. 1990).

The failure of the relation,
i.e. the existence of large background LSBGs (such as the
serendipitously discovered Malin 1; Bothun et al.\ 1987), or of a
population of high surface brightness (compact) dwarfs in a cluster
(Drinkwater \& Gregg 1998), or indeed of a population of galaxies so compact 
as to masquerade as stars and, hence, be missed
altogether from galaxy samples (e.g. Arp 1965; Drinkwater et al. 2000b; Phillipps et al. 
2001) could have a dramatic effect on our
perception of the galaxy population as a whole (see Impey \& Bothun 1997).

The primary factor preventing a resolution of the controversy has been
the non-existence of suitable samples of confirmed dwarfs of known distance
(and hence luminosity) with well defined selection limits. Generally, 
dwarf galaxy 
samples have been either subjectively (for instance, morphologically)
selected sets of cluster galaxies (e.g. Ferguson 1989) or objectively
chosen sets of galaxies with unconfirmed cluster membership/unknown
distances (e.g. Irwin et al. 1990).

However, the development of a new generation of multi-object spectrographs,
exemplified by the `2 degree field', or 2dF, multi-fibre spectrograph
on the Anglo-Australian Telescope (AAT) (see e.g. Taylor, Canon \& 
 Parker 1998), 
 has  made it possible to conduct a truly complete
spectroscopic survey of a given area on the sky, down to well
determined, faint limits, irrespective of image morphology or any
other pre-selection of target type, and thus obtain redshifts for 
a large and unbiased sample of targets (Phillipps 1997).
Our  Fornax Cluster Spectroscopic Survey (FCSS) aims to  exploit
the huge multiplexing advantage of 2dF in this way by surveying, in total, a region of 
  12~deg$^{2}$ centred on the Fornax Cluster of galaxies. 
The prime reason for choosing this area of  sky area 
was, of course, the presence of the Fornax Cluster itself.
  Fornax, with Virgo, is the nearest reasonably rich cluster
(approximately Abell Richness Class 0 -- it is supplementary cluster
S0373 in Abell, Corwin \& Olowin 1989).

The targets of the FCSS 
 encompass both cluster galaxies, of a wide range of types and
magnitudes, and background galaxies (over a similarly
wide range of morphologies), as well as Galactic stars and QSOs
 (Drinkwater et al. 2000a). 
  Our main motivation for an all-object survey was to
determine cluster membership for as complete a sample of objects as possible
across a wide range of surface brightnesses. We wished
to test, in particular, whether the usual assignment of LSBGs to the
cluster and high (or even `normal') surface brightness faint
galaxies to the background is justified. The situation regarding HSBGs
is even less clear than for LSBGs -- the question of the very existence
of normal or high surface brightness dwarf galaxies (excepting
possibly pathological cases like M32) has been highly controversial -- see,
for example, the contrasting views expressed in Ferguson \& Sandage
(1988) and Irwin et al.\ (1990). An up to date discussion is given in 
Trentham \& Hodgkin (2002).

  Drinkwater et al. (2000a) (hereafter Paper~I) outlined the aims 
 and methods of the FCSS and presented the results from the 
 first  2\degr\, diameter field surveyed. Drinkwater et al. (1999) 
 (Paper~II)  described the nature of several compact galaxies found beyond 
 the cluster. Meyer et al. (2001) ( Paper~IV) discussed the quasars 
 found in the FCSS data. Drinkwater et al. (2000b) (Paper~III) 
 and Phillipps et al. (2001) (Paper~V) described the discovery of 
 5 extremely compact, high surface brightness dwarf galaxies 
 within the cluster. These objects (dubbed 
 `Ultra-Compact Dwarfs' -UCDs) have intrinsic sizes $\la$100~pc. They are 
 more compact and significantly less luminous than other known compact 
 dwarf galaxies, yet are much brighter than globular clusters.

The data from the first field (see Paper~I) revealed that 
 the Fornax Cluster is clearly defined kinematically. 
  A total of 24 galaxies from the data lie within the 
 cluster (this includes the 5 UCDs described in Papers~III and V). 
  This is, to our knowledge, the first complete sample of dwarf galaxies of known 
 distance (and hence luminosity) with well defined selection limits. 
  In this paper we study the nature of the objects in this complete sample.  
    Section 2 briefly describes the methods and present status of the FCSS.  
 Section~3 describes the contents of the complete Fornax Cluster sample and 
 presents  $B$ and $R$ photometry,   
 radial intensity profiles and radial colour distributions for  
  all 24 confirmed cluster members. Using this data we are able to explore (in Section 4) 
 the distribution in luminosity and surface brightness 
 of the confirmed cluster members. Section 5 presents 
 our conclusions.

\section{The Fornax Cluster Spectroscopic Survey}

The FCSS survey-area will ultimately comprise  four separate 
2dF fields, covering the majority of the Fornax Cluster area
 (see fig.~1 and table~1 of Paper~I). The results presented 
 in this paper are derived from the first of these four fields to be completed. 
 Field~1 is centred on the large galaxy NGC~1399 at the 
 centre of the cluster (R.A.=03$^{h}$38$^{m}$29.0$^{s}$, 
  Dec.=35\degr27\arcmin01\arcsec, J2000). 

In common with the main 2dF Galaxy Redshift Survey 
(2dFGRS; see Colless et al. 2001 and references therein), we
have chosen to select our targets from catalogues based on UKST Sky
Survey plates digitised by the Automatic Plate Measuring (APM) facility
 at Cambridge (see Irwin, Maddox \& McMahon 1994). However, 
unlike other galaxy surveys, which only
select resolved images for spectroscopic measurement, we avoid any
morphological pre-selection and include all objects, both resolved and
unresolved (i.e.\ `stars' and `galaxies').  This means that we can
include galaxies with the greatest possible range of surface
brightnesses: our only selection criterion is the (blue) magnitude
limit. Including objects normally classified as stars greatly
increases the size of our sample but it is the only way to ensure
completeness. 

In order both to cover
a large number of targets and to go significantly deeper than previous
spectroscopic cluster surveys we chose to limit our survey at a $b_{j}$ magnitude
of 19.7. This is marginally deeper than the main 
2dFGRS, which is limited at $b_{j} = 19.45$.
 $b_{j}$ is the natural $B$ band photographic system of the UKST data. 
  This is related to the standard Cousins $B$ magnitude by 
 $b_j=B-0.28\times(B-V)$ (Blair 
 \& Gilmore 1982). 
Targets are also chosen to be fainter than $b_{j} = 16.5$, since the
photometry of brighter objects is problematic with the UKST/APM data
(particularly because of image saturation).

Note that the selection limits of the overall APM
 object  catalogue (images larger than a certain area at a given isophotal detection
threshold) do not impinge on our target selection, since any galaxies
brighter than $b_{j}$ = 19.7 and with surface brightness high enough
to be observable with 2dF (see below) are well above the APM limit
(see also Paper I).

Our input catalogue for the  FCSS is a standard APM `catalogues' file
(Irwin et al. 1994) of field F358 from the UKST 
Southern Sky Survey. The field is
centred at R.A.=03$^{h}$37$^{m}$55\fs9, Dec.=--34\degr50\arcmin14\arcsec\ (J2000) and
the region scanned for the catalogue file is 5.8$\times$5.8~deg$^{2}$,
 approximately centered on the Fornax Cluster. The APM image catalogue lists image positions,
magnitudes and morphological classifications (as `star', `galaxy',
`noise', or `merged') measured from both the blue and red survey
plates.  The `merged' image classification indicates two overlapping
images: at the magnitudes of interest for this project the merged
objects nearly always consisted of a star overlapping a much fainter
galaxy. All the positions are measured from the more recent red survey
plate (epoch 1991 September 13 compared to 1976 November 18 for the
blue plate) to minimise problems with proper motions.  The APM
catalogue magnitudes are calibrated for unresolved (stellar) objects
only (see Bunclark \& Irwin 1984), so we supplemented these with total magnitudes for the galaxies
measured by direct analysis of the digitised
plate data. These magnitudes were
obtained by fitting exponentials to the APM galaxy surface brightness
profiles (see Phillipps et al. 1987 and Morshidi-Esslinger, Davies \& Smith 1999)
across the surface brightness range 22.7 to 25.7~B~mag~arcsec$^{-2}$ (i.e.
between the possibly saturated inner parts of the images and the
sky noise dominated outer parts). Central surface brightnesses and 
exponential scale lengths could be determined in this way and
combined to give pseudo-total magnitudes (see
Paper I for further details); we refine
these measurements for the cluster objects in the present paper.

Our target selection then consisted simply of taking all objects from the
APM catalogue in each of our four 2dF fields with magnitudes in the
range $16.5 \leq b_j \leq 19.7$. 
 We take the distance to the
cluster to be 20~Mpc (a distance modulus of 31.5~mag) as derived
by Drinkwater, Gregg \& Colless (2001b). This leads to an absolute
magnitude range for  cluster galaxies --14.8 $\la$ $M_{B}$ $\la$ --11.6, so
all the cluster galaxies from FCSS are dwarfs in terms of their luminosity. 
 We did not apply any morphological selection,
although the APM image classifications from the blue survey plate were
used to determine which photometry to use (i.e. the catalogued APM
magnitude in the case of unresolved images, or our fitted magnitudes for
resolved images).

Although our targets cover a wide range of surface brightnesses (see fig.~1
 of Paper I),
our final spectroscopic sample inevitably suffers from
incompleteness at the low surface brightness end. It is not
possible to measure spectra for the faintest LSBGs catalogued
 (at central surface brightness around 24.5~B~mag~arcsec$^{-2}$) in
reasonable exposure times even though the multiplex advantage of 2dF
enables us to profitably expose for longer times and hence
go fainter than most previous nearby cluster work. Two hour
exposures enabled us to obtain measurable spectra for
galaxies with central surface
brightnesses down to 23.2~B~mag~arcsec$^{-2}$
with a success rate of $\geq$80 per cent (see Paper I).  In
order to try to extend our surface brightness limit, we have also
experimented with much longer exposures (around 4 hours) for the
lowest surface brightness subset and have obtained successful
spectra for some objects with surface brightnesses as faint as
23.7~B~mag~arcsec$^{-2}$. This forms the practical lower limit to our cluster
galaxy sample.

We observed all our targets with an identical observing setup for 2dF:
the 300B grating and a central wavelength setting of 5800\AA\ giving a
wavelength coverage of 3600--8010\AA\ at a resolution of 9\AA\ (a
dispersion of 4.3\AA\ per pixel).

The 2dF facility includes its own data reduction package, {\sc 2dfdr},
 which permits fast, semi-automatic reduction of data direct from the
instrument. However, when we started the  FCSS project 
{\sc 2dFdr} was still under
development, so we chose instead to reduce the data with the {\sc
DOFIBERS} package in {\sc IRAF}\footnote{{\sc IRAF}  is distributed by the National
Optical Astronomy Observatories, operated by the Association of
Universities for Research in Astronomy, Inc.  under cooperative
agreement with the NSF.}, as described in Paper I. 
We have since reduced the data independently
with {\sc 2dFdr}, obtaining entirely comparable results (Deady 2002).
In particular, 98 per cent of spectra gave the same velocity shift (see below)
to within the quoted errors; the remaining 2 per cent turned out, on further inspection, to have
spurious features which had not been removed by the cleaning of the original
data.

\begin{figure}
  \epsfig{file=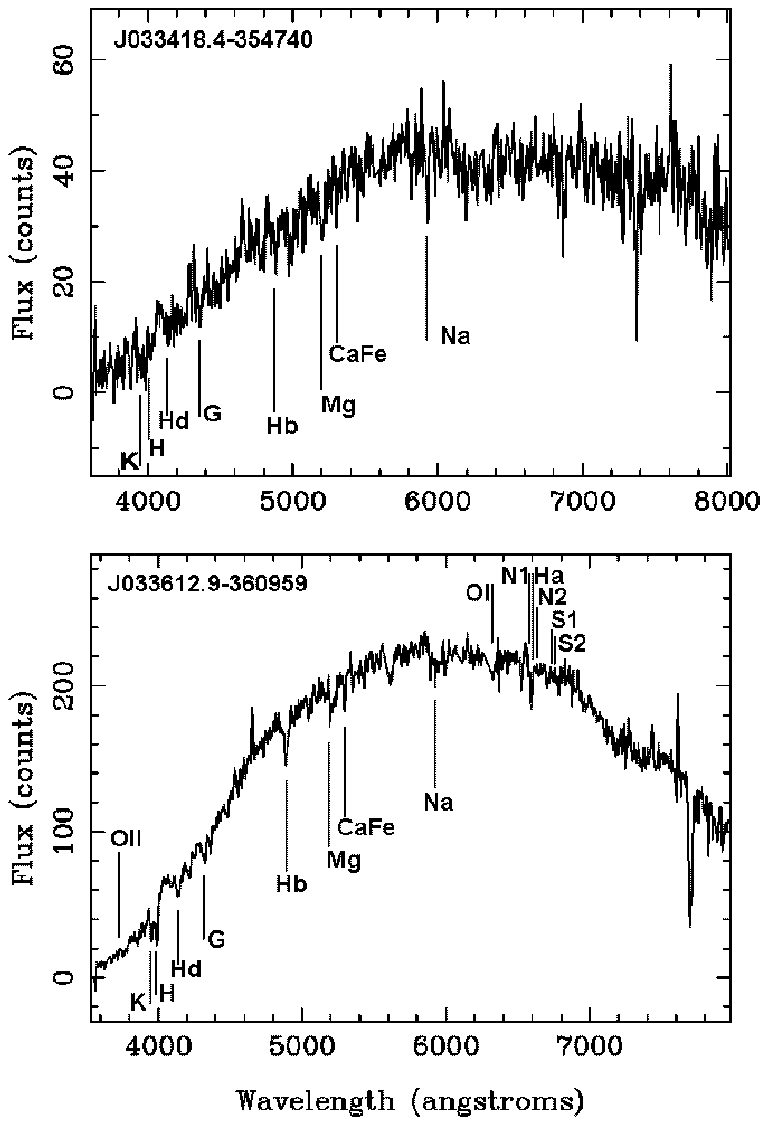}
  \caption{The unfluxed 2dF spectra for 2 of the Fornax Cluster sample. 
 The full set of spectra can be viewed at http://www.star.bris.ac.uk/pjb/2dfspectra.htm.}
\end{figure}

Redshifts are determined in a standard fashion
via cross-correlation of the spectra with templates,
using {\sc RVSAO} in {\sc IRAF} (Kurtz \& Mink 1998). However, as we required
an automated system capable of coping with a large number of stars as well
as galaxies, we have used stellar template spectra of various types, rather
than absorption line galaxy spectra. These were chosen from the library of
Jacoby, Hunter \& Christian (1984). In addition we use emission line
galaxy and quasar templates. For each spectrum we determined the 
 best-matching template, the Tonry-Davies (1979) R coefficient, 
 the redshift and its error. 
(Again see Paper I for more details and illustrations of the template
spectra used).
  The redshifts are measured as radial velocities in units of
$cz$ and are subsequently converted to heliocentric values.  By
choosing the template giving the best $R$ coefficient we can determine
not only the redshift, but a first estimate of the object type. We
only accept identifications with $R\ge3$.  Objects with redshifts of
$\approx$500~km\,s$^{-1}$ or less are Galactic stars for which the best
template indicates the stellar spectral type.  At higher redshifts,
external galaxies are separated into absorption-line types if they
match one of the stellar spectra or emission-line types if they match
the emission-line galaxy template best.

Typical velocity errors are around 64~km~s$^{-1}$, while comparison with 44
galaxies with redshifts given in the NASA/IPAC Extragalactic Database (NED)
reveals a negligible external mean error of $7 \pm 17$ km~s$^{-1}$ and
rms scatter of 110~km~s$^{-1}$, consistent with our internal errors given
the heterogeneous data compiled in NED.

\begin{table*}
\begin{minipage}{140mm}
\caption{The confirmed members of the Fornax cluster sample}
\begin{tabular}{lrrrllrr} \hline
Name &     &     &    &       & FCC  & $cz \; \; \; \;$      & \\
FCSS & FCC & FLSB & NG & Other & Type & km~s$^{-1}$ & Template \\ \hline
J033418.4-354740 & 132     & --  & -- &      & dE2      & $1883 \pm 98$ & M0\\
J033612.9-360959 & 164     & 64  & -- &      & dS0(5),N & $1485 \pm 45$ & K5\\
J033653.3-345618 & 181     & 71  & 56 & HPW19 & dE2,N   & $1333 \pm 69$ & G6\\
J033703.3-353804 & --      & --  & -- & UCD1 &          & $1491 \pm 39$ & K5\\
J033718.0-354157 & 194     & 252 & -- &      & dE3      & $1237 \pm 84$ & G0\\
J033723.3-345400 & 195     & 80  & 55 & H14  & dE5,N    & $1315 \pm 69$ & G6\\
J033734.1-354945 & 196     & 83  & 23 &      & dE6      & $1797\pm 129$ & M0\\
J033754.7-345255 & 200     & 256 & -- &      & dE2     & $1184 \pm 110$ & F6\\
J033806.3-352858 &  --     & --  & -- & UCD2 &          & $1312 \pm 57$ & K5\\
J033816.7-353027 & (B1241) & 95  & -- &      & (dE3,pec?) & $2012 \pm 91$ & G6\\
J033821.5-351535 & 211     & --  & 71 &      & dE2,N    & $2192 \pm 49$ & F6\\
J033854.1-353333 & --      & --  & -- & UCD3 = CGF1-4 & & $1591 \pm 36$ & K5\\
J033905.8-360556 & 221     & 108 & 43 &       & dE4,N   & $1724 \pm 77$ & F6\\
J033919.5-354334 & 223     & 114 &  6 & HPW28 & dE0,N   &  $902 \pm 83$ & K5\\
J033935.9-352824 & --      & --  & -- & UCD4 = CGF5-4&  & $1920 \pm 40$ & K5\\
J033952.5-350424 & --      & --  & -- & UCD5  &         & $1321 \pm 66$ & F6\\
J034001.3-344529 & 230     & 120 & 76 & HPW29 & dE5,N   & $1149 \pm 84$ & G0\\
J034023.5-351636 & 241     & 124 &  5 & HPW30 & dE0,N  & $2045 \pm 107$ & F6\\
J034033.8-350122 & 245     & 126 &  3 &       & dE0,N   & $2265 \pm 43$ & K5\\
J034042.3-353940 & 247     & 127 & 40 &       & dE3/Im? & $1097 \pm 108$ & F6\\
J034100.8-354433 & 254     & 129 & 22 &       & dE0,N   & $1517 \pm 94$ & G0\\
J034131.8-353521 & 264     & 135 & -- &       & dS0(8),N & $2033 \pm 64$ & G6\\
J034159.4-352053 & (B1554) & --  & -- & CGF3-1 & (E)    & $1667 \pm 59$ & K5\\
J034217.3-353226 & 274     & 141 & 15 & HPW39 & dE0,N   &  $977 \pm 85$ & K5\\ \hline
\end{tabular}
\end{minipage}
\end{table*}

In the second stage of the identification process we check each
identification interactively (cf. Colless et al. 2001),
using the {\sc RVSAO} package to
display the best cross-correlation and the object spectrum
with common spectral features plotted at the calculated redshift. In the
small number of cases
when the redshift is obviously wrong (e.g.\ with the calcium {\sc H} and 
   {\sc K}
lines clearly present but misidentified), it is flagged as being incorrect
or in some cases is recalculated interactively.
Objects still not dealt with successfully at this stage are flagged to be
re-observed.

Note that the 
2dF spectra, though of low resolution and unfluxed, are still useful
for more detailed analysis than simple redshift measurements and
object classifications.  However, we 
defer any detailed analysis of the spectra in terms of, e.g.,
emission or absorption line equivalent widths, line ratios, 
absorption line indices or rough metalicities
to later papers.

\section{The Fornax Cluster Sample}

\subsection{Composition of the sample}

Within Field 1, we found 24 galaxies with redshifts which place them in the
Fornax Cluster (mean $cz \simeq 1450$ km~s$^{-1}$).
(In fact, 26 of the input APM catalogue objects have such  
velocities, but closer inspection reveals one to be an individual HII
region of the large irregular galaxy NGC~1427A and the other to be
a mistakenly demerged part of the image of the giant galaxy NGC~1381.)
Assignment to the cluster is unambiguous, as the cluster is well
isolated in redshift space (see Paper I). Our magnitude
selection limits place all the sample objects between roughly
$M_{B}$=--14.8 and $M_{B}$=--11.6; more precise magnitudes are determined
in Section~3.2 below. Five of the objects were originally observed as `stars', i.e.
they had unresolved or only marginally resolved images on the plate material
used for the APM catalogue.
The relatively small number of cluster galaxies is largely due to the 
lower bound on the surface brightness at which successful fibre spectra
can be obtained. Many dwarfs at these faint magnitudes are expected to
have surface brightnesses well below our practical 2dF limit at
around 23.7~B~mag~arcsec$^{-2}$ (see, e.g., Irwin et al. 1990, Kambas et al. 2000).

Our 24 objects are listed in R.A. order in Table 1. We give their
official IAU designations along with their names in other recent Fornax catalogues.
FCC refers to Ferguson's (1989) Fornax Cluster Catalog, FLSBG numbers
are from the Fornax Low Surface Brightness Galaxy catalogue of Davies et al.
(1988), continued in Irwin et al. (1990) and the UCDs are from Phillipps
et al. (2001). NG refers to the galaxies listed by Caldwell (1987).  
Note that the two non-compact galaxies without FCC numbers were given
background galaxy numbers by Ferguson (FCCB 1241 and 1554), but clearly
these designations are now inappropriate (see Section 4.2 for a full 
 discussion of this issue). The latter is also CGF3-1 from Hilker et al.
(1999), and UCD3 and UCD4 are CGF1-4 and CGF5-4 from the same source.
  Finally
objects FCC 181, 195, 223, 230, 241 and 274 were earlier studied by Hodge
(1960) or Hodge, Pyper \& Webb (1965), where they were the dwarfs
HPW 19, H 14 and HPW 28, 29, 30 and 39, respectively (see Irwin et al. 1990).

As noted, 17 of the 19 non-compact galaxies appear as cluster galaxies
in Ferguson's FCC,
and we have included Ferguson's assessment of their morphological types
from his large plate scale Las Campanas plates. All are classified dE
or, in two cases, dS0; 12 were judged `definite members',  4 as `probable members'
  and 1 as a `possible member' (see Section 4.2).

The remaining columns in Table 1 give our derived heliocentric redshift
(given as a `velocity', $cz$) and the best fitting stellar template (there
are no emission line objects in the sample once the NGC~1427A
HII region has been removed).
Example spectra are shown in Fig.~1 (all of them
can be accessed at http://www.star.bris.ac.uk/pjb/2dfspectra.htm). Remember that no correction
for throughput as a function of wavelength in each fibre has been attempted,
so the continuum shape in the plots is fairly arbitrary.

The majority (17) of the spectra are best matched by G (7) or K (10)
star templates, with 6 matches to F types and 2 to M types. The fact that
none of our objects displays any sign of star formation  corroborates
Ferguson's dE and dS0 classifications. 
 The non-detection of any dwarf irregular (dI) galaxies within Field 1 is noteworthy.
 Field 1 is centered on NGC1399 and covers the central part of the cluster. 
 Drinkwater et al. (2001a) obtained Flair-II  
 spectra for 108 brighter ($b_j<$18) members of the cluster over a 
 6\degr\, field.  They found that among the dwarfs in 
 their sample ($M_{B}\ge$--16.5),  the early-type dwarfs dominate in the cluster 
 centre (within a  radius of 100 arcmin) with late-type dwarfs being concentrated 
 in the outer regions of the cluster.

  Recent theoretical work on dynamical evolution  
 has suggested that, in rich clusters,  
    tidal forces from  the cluster's tidal field 
  can result in significant morphological evolution, enough 
 to convert dIs to dEs  (Moore, Lake \& Katz 1998).  Boyce et al. (2001) 
 suggested that such  forces may explain their observation that the core of 
 the rich cluster Abell~868 is devoid of dIs but that there are similar proportions 
 of dIs to dEs beyond the cluster centre.  Mayer et al. (2001) have suggested 
 that a related  process (`tidal stirring') may explain the distribution of dI and dE 
 galaxies in the Local Group. In this process  tidal interactions between 
 the dwarfs  and the larger member galaxies cause the tidal instabilities which induce 
 evolution from dI to dE.  In a poor, but centrally concentrated  cluster 
 like Fornax, both of these processes may  be at work.

\subsection{Photometry of the sample}

\begin{table*}
\begin{minipage}{100mm}
\caption{Derived photometric parameters for the Fornax cluster sample}
\begin{tabular}{lrcrrrr}  \hline
Name & $\alpha \; \;$ & $\mu_{0}$ & $B$ & $R$ & $M_B$ & $a \;$ \\
FCSS & arcsec   & B mags arcsec$^{-2}$    &     &     &       & kpc \\  \hline
J033418.4-354740 & 3.64      & 23.2   & 18.8 &  17.3 & --12.7  & 360 \\
J033612.9-360959 & 5.93      & 21.9   & 16.7 &  15.3 & --14.9  & 580 \\
J033653.3-345618 & 5.94      & 22.9   & 17.6 &  16.3 & --13.9  & 580 \\
J033703.3-353804$^{\star}$ & 0.89      & 21.5   & 20.2 &  18.7 & --11.4  & $<100$ \\ 
J033718.0-354157 & 4.13      & 23.3   & 18.6 &  17.3 & --12.9  & 410 \\
J033723.3-345400 & 7.27      & 22.8   & 17.1 &  15.7 & --14.4  & 720 \\
J033734.1-354945 & 7.90      & 23.7   & 18.2 &  16.5 & --13.3  & 790 \\
J033754.7-345255 & 4.62      & 23.0   & 17.8 &  16.4 & --13.7  & 450 \\
J033806.3-352858$^{\star}$ & 1.41      & 22.3   & 19.9 &  18.3 & --12.4  & $<140$ \\
J033816.7-353027 & 5.43      & 22.7   & 17.5 &  16.1 & --14.0  & 570 \\
J033821.5-351535 & 4.35      & 21.5   & 16.5 &  15.1 & --15.0  & 470 \\
J033854.1-353333$^{\star}$ & 0.98      & 20.7   & 18.9 &  17.2 & --13.6  & $<100$ \\ 
J033905.8-360556 & 3.86      & 22.6   & 18.0 &  16.5 & --13.5  & 380 \\
J033919.5-354334 & 9.21      & 23.2   & 16.6 &  15.3 & --14.9  & 900 \\
J033935.9-352824$^{\star}$ & 1.00      & 21.9   & 19.9 &  18.5 & --12.5  & $<100$ \\
J033952.5-350424$^{\star}$ & 0.97      & 21.8   & 20.1 &  18.7 & --11.6  & $<100$ \\
J034001.3-344529 & 5.30      & 22.6   & 17.6 &  16.4 & --13.9  & 520 \\
J034023.5-351636 & 8.76      & 23.7   & 17.1 &  16.0 & --14.4  & 860 \\
J034033.8-350122 & 6.85      & 22.3   & 16.3 &  15.0 & --15.2  & 670 \\
J034042.3-353940 & 6.02      & 23.4   & 18.1 &  17.1 & --13.4  & 590 \\
J034100.8-354433 & 9.42      & 24.6   & 18.1 &  16.6 & --13.4  & 920 \\
J034131.8-353521 & 6.08      & 21.9   & 17.0 &  15.7 & --14.5  & 600 \\
J034159.4-352053 & 3.23      & 22.1   & 17.8 &  16.2 & --13.8  & 320 \\
J034217.3-353226 & 6.30      & 22.8   & 16.9 &  15.8 & --14.7  & 620 \\  \hline
\end{tabular}
\end{minipage}
\end{table*}

As noted earlier, initial photometry for all the catalogue galaxies was
derived from the APM scan data in automated fashion, assuming pure
exponential radial profiles and circular isophotes (cf. Morshidi-Esslinger
et al. 1999). We have now improved on this photometry 
 by using 
  photographic data  from a IIIaJ $B$ plate and a Tech Pan $R$ film
from the UKST, digitised by the SuperCOSMOS measuring machine at the
Royal Observatory, Edinburgh (e.g. Hambly, Irwin \& MacGillivray 2001).
 Table~2 presents the results of this photometry. 
  The
data were calibrated (Deady 2002)
onto the standard Cousins $B$ and $R$ 
   system via comparison of the surface brightness profiles of four suitable, lowish
surface brightness dwarfs (so as to avoid problems of saturation at
moderate to high surface brightness on the plates) with corresponding
CCD data from Davies, Phillipps \& Disney (1990). 
The methodology was as in Phillipps \& Parker (1993).
Note that none of the calibrating galaxies are actually in the present cluster
galaxy sample, either being outside the 2dF area or of too low a surface
brightness for 2dF spectroscopy. The Davies et al. (1990) calibration is
closely similar  (to 0.1~mag) to both that of the original APM low
surface brightness galaxy survey of Davies et al. (1988) and the Fornax
Cluster Catalog of Ferguson (1989). After independently calibrated 
Curtis Schmidt data became available (Karick et al., in prep.)
we were able to confirm that our photographic and CCD $B$ magnitude
scales were in excellent agreement, too.

To determine total magnitudes for our galaxies, we used the {\sc SEXTRACTOR} package
(Bertin \& Arnouts 1996) to generate Kron magnitudes (Kron 1978). We selected
an aperture size of 3.5 Kron radii (cf. Metcalfe et al. 1995),
which contains 99.3 per cent of the light for an exponential profile. Our
magnitudes should therefore be total to within $\simeq$0.01~mag. A
more significant source of error will be the determination of the
background sky level. A realistic measure of the uncertainty in this
was judged to be $\sigma_{sky}/10$, where $\sigma_{sky}$ is the width
(standard deviation) of the histogram of local sky values. Repeating
the Kron radius and magnitude calculation with the background varied
by this amount gives our estimate of the magnitude errors from this
source. Allowing also for possible systematic zero point errors of $\simeq$0.05~mag
 we find final photographic magnitude errors $\simeq$0.1--0.2~mag.
The $B$ and $R$ magnitudes for all our cluster galaxies are summarized in columns 
 4 and 5 of Table~2. Note, however, that the photographic data 
   of  the UCDs (denoted 
 by an asterix in Table~2) suffer
 from saturation, so the $B$ and $R$ values in Table~2 only provide 
  faint magnitude limits for these objects. 
 Column 6 presents $M_{B}$ values assuming 
 D=20~Mpc (see Section 2). In this column, we have avoided the problems of 
 the saturation of the photographic data for the UCDs, 
 by  including, for these objects, the absolute magnitudes 
  derived from  Curtis Schmidt CCD data by Phillipps et al. (2001), and corrected from $b_{j}$ 
 to $B$ by assuming $B-V$$\simeq$0.7.

The $B-R$ colours are all rather similar (nearly all between 1.2 and 1.5), 
and are consistent with the spectral
classifications as quiescent dE or dS0 galaxies. Aside from the UCDs, the 
  galaxies have a mean 
colour of $<B-R>$=1.33$\pm$0.03, essentially identical
to the $<B-R>$=1.38 reported by Davies et al. (1990) for
their (presumed) cluster LSBGs. Although the numbers are small,
there is a (comforting) slight trend for redder colours for the objects
which match best to later type stellar spectra (in particular
the two M0 matches have an average $B-R$ of 1.54).

\begin{figure*}
  \epsfig{file=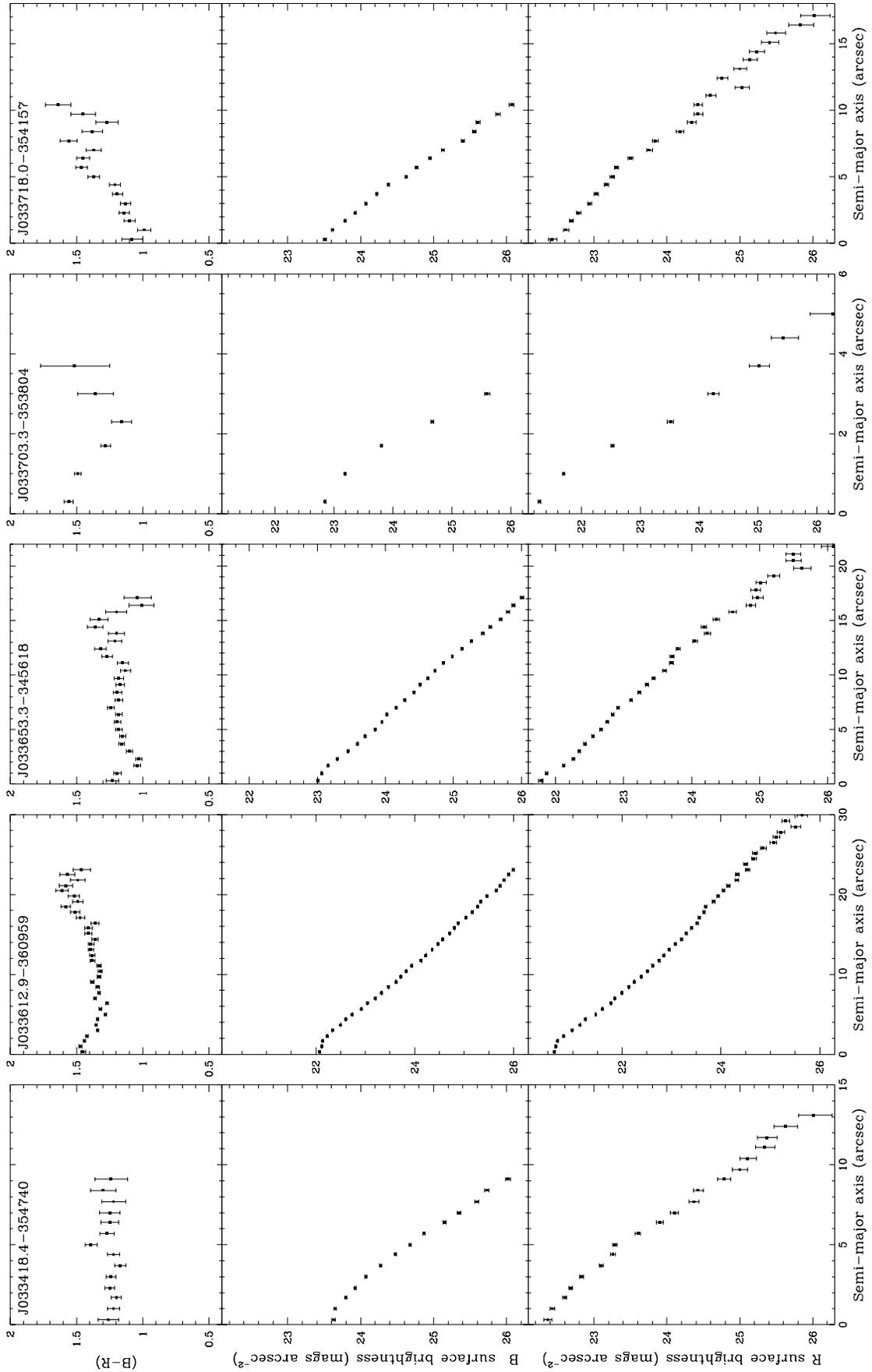}
  \caption{Radial intensity profiles and radial colour profiles 
 for the Fornax Cluster sample}
\end{figure*}

\addtocounter{figure}{-1}
\begin{figure*}
   \epsfig{file=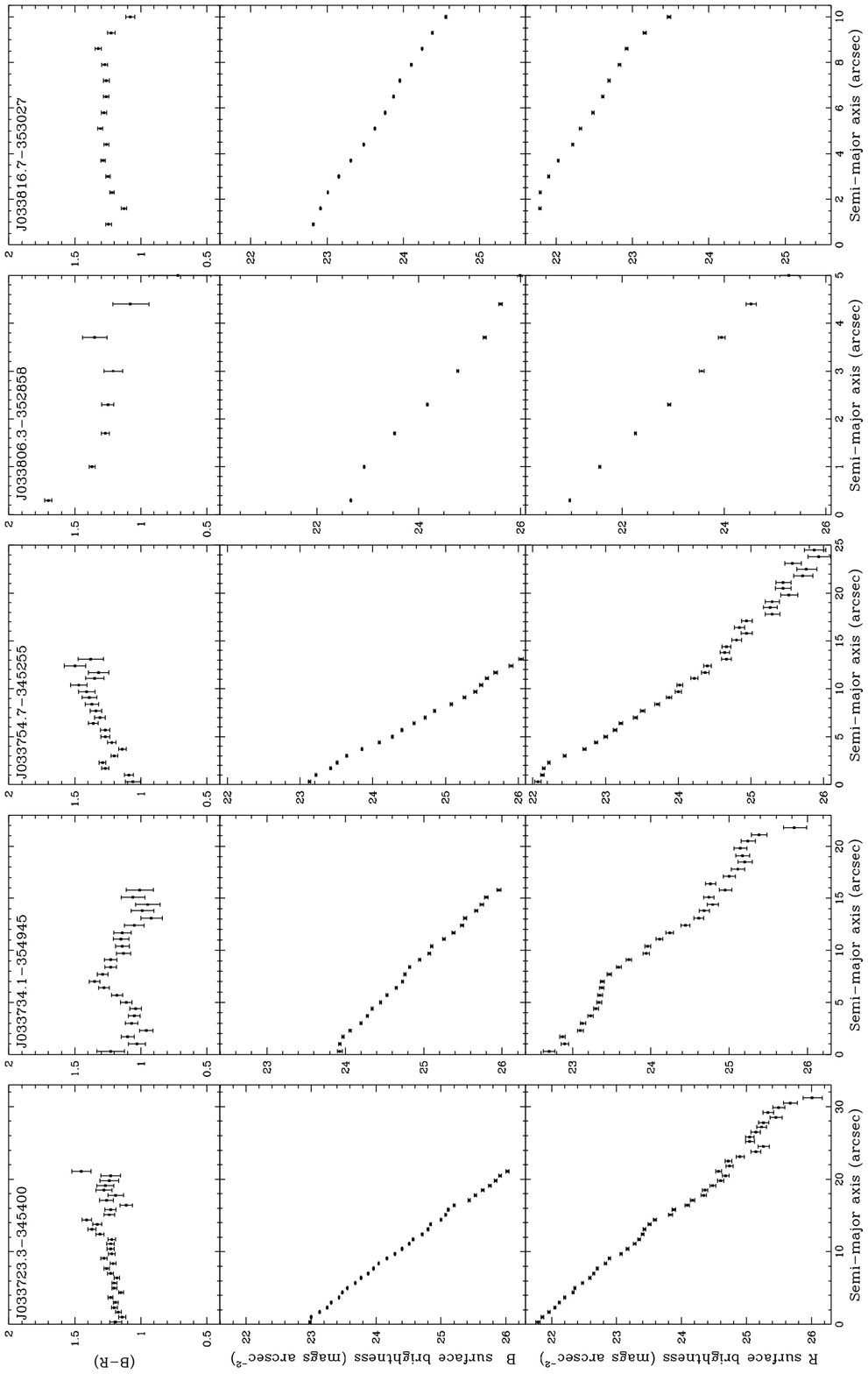}
  \caption{continued}
\end{figure*}

\addtocounter{figure}{-1}
\begin{figure*}
  \epsfig{file=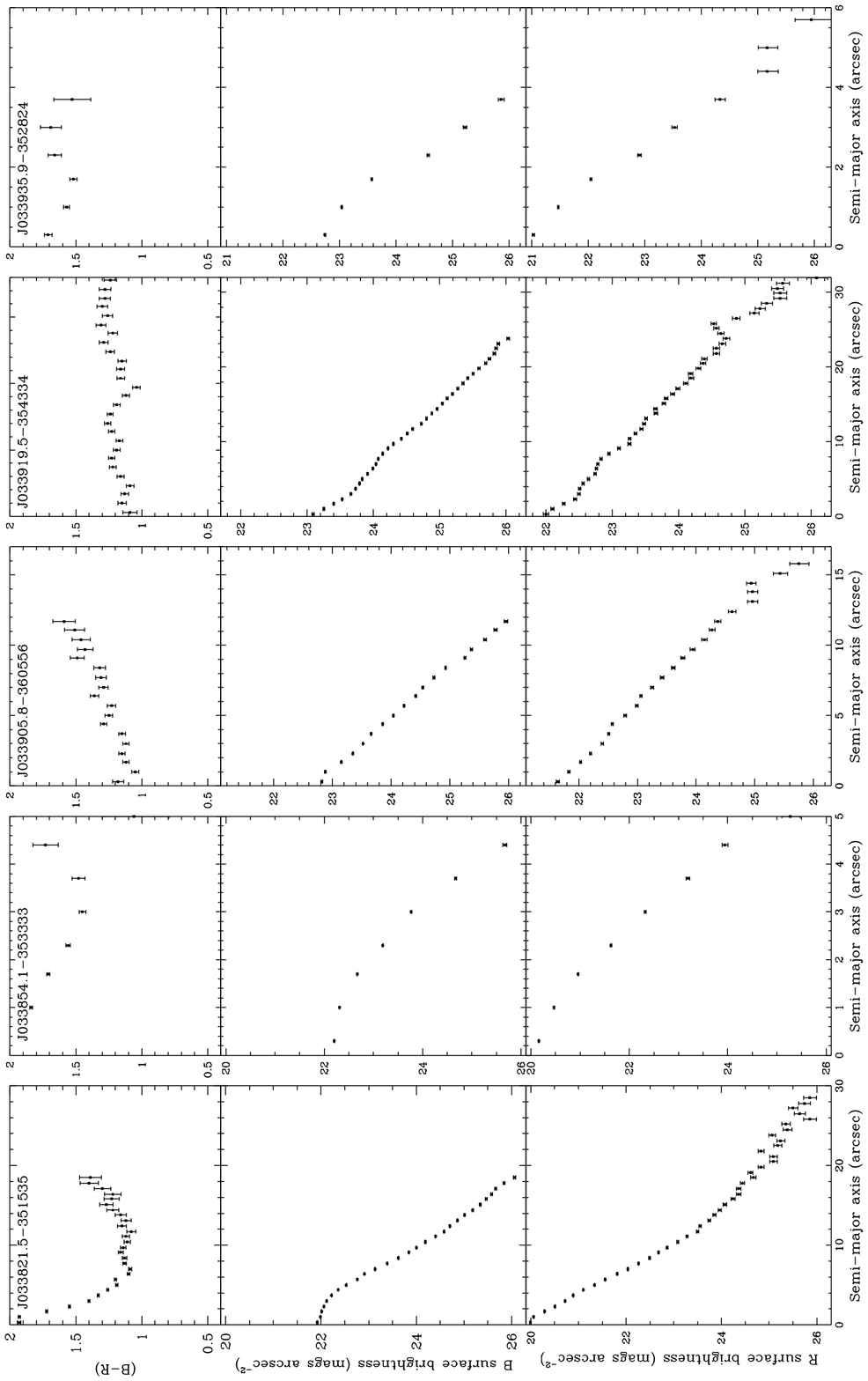}
  \caption{continued}
\end{figure*}

\addtocounter{figure}{-1}
\begin{figure*}
  \epsfig{file=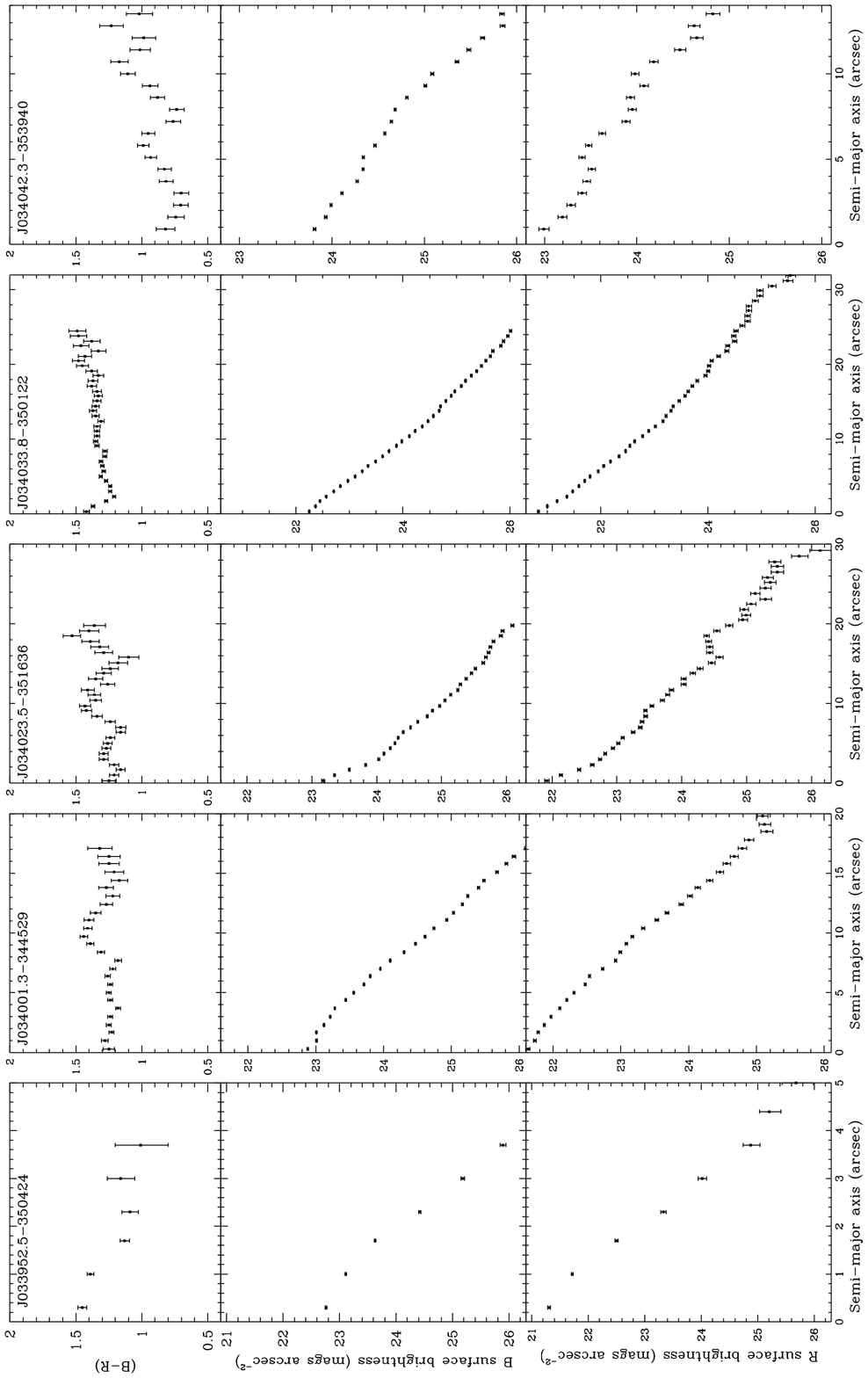}
  \caption{continued}
\end{figure*}

\addtocounter{figure}{-1}
\begin{figure*}
  \epsfig{file=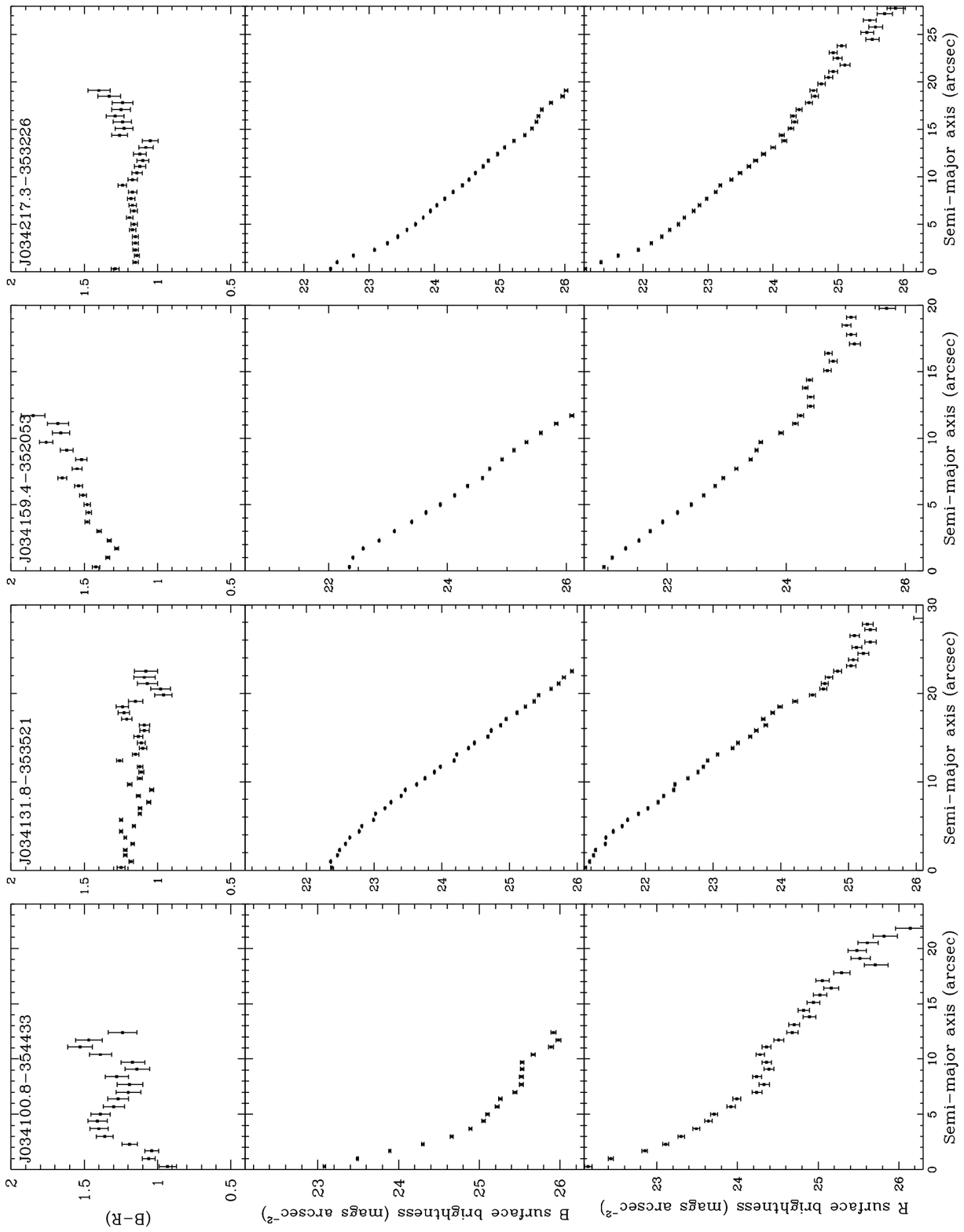}
  \caption{continued}
\end{figure*}

The {\sc STARLINK} routine {\sc ELPROF}  was used to determine
the radial intensity profile, as a function of semi-major axis length,
for the cluster dwarfs (as well as the calibrating galaxies) from the
photographic $B$ and $R$ data. The radial intensity 
 profiles and radial colour distributions 
 are illustrated in 
Fig.~2. Most of the radial intensity profiles  are well fitted by a simple exponential law (straight
line in the plot), typical of dwarf galaxies, though some do show some
curvature. In some cases this is due merely to the saturation of the
$B$ band measurements at $\simeq$22~B~mag~arcsec$^{-2}$. This is less of a problem
with the $R$ profiles as the Tech Pan films have somewhat better
dynamic range. For completeness we have therefore also fitted the
more general Sersic (1968) law (cf. Davies et al. 1988). This
fitting confirms that 13 of the 19 normal (i.e. not ultra-compact)
dwarfs have best fitting indices $n = 1.0 \pm 0.2$ (in the $B$ band),
where $n = 1$ corresponds to a pure exponential. Four of the remainder have
$n$ between 0.4 and 0.8 (a pure de Vaucouleurs (1959) law, appropriate for
giant ellipticals, would have $n = 0.25$), while two have $n \simeq 1.4$.
Given that the differences in $n$ calculated for the $B$ and $R$ band
profiles are typically around 0.2 (even when there is no systematic
colour gradient), that is probably a reasonable estimate of the level of
uncertainty in $n$. For subsequent fitting we thus assume that the
standard exponential profile is a sufficiently good fit in all cases.

Columns 2 and 3 of Table 2 therefore contain 
 the scale sizes $\alpha$ (in arcsec) and extrapolated 
  central surface brightnesses $\mu_{0}$ of the exponential fit to the $B$ band
 radial intensity profiles. Column 7 lists the scale size in kpc (assuming D=20~Mpc). 
 Fitting was carried out in the range between
22.5 and 26~B~mag~arcsec$^{-2}$. For the compact objects, the tabulated values 
 are merely limits on the intrinsic surface brightnesses and scale lengths 
  due to the  effects of the point-spread function  (as discussed in more
detail in Paper V). Errors in the scale sizes from the fitting
procedure are typically of order 10 per cent, and the differences between the
$B$ and $R$ scale lengths are at the same level in almost all cases
(though a couple show systematic colour gradients and therefore
differences up to $\sim$20 per cent). One galaxy, FCSS J034100.8-354433,
is difficult to treat as 
simply as the others as it appears to have a two component structure.
In the table we give the parameters for its `outer disc' (which has a
very low surface brightness around 24.5~B~mag~arcsec$^{-2}$, 
making it an outlier in the plots below),
but it also has a much more steeply rising `inner bulge' with actual central
surface brightness more like 23~B~mag~arcsec$^{-2}$ (hence the  
 successful spectroscopy of this object), 
  and a scale size of around 2~arcsec.
Given the small range of colours for most of the sample, unsurprisingly no
correlation between colour and surface brightness or scale length is
apparent.

\section{Discussion}

\subsection{The Bivariate Brightness Distribution (BBD) of  the cluster sample}

 Two  necessary steps towards  any full understanding of the processes 
 of galaxy formation and evolution are (i) the compilation of a 
 full and complete census of the local extragalactic 
 population and (ii) an  accurate and detailed 
 quantification of this  population. Traditionally 
 astronomers have sought to quantify the local galaxy population 
 by constructing galaxy luminosity functions (LFs). 
 The LF of a sample of galaxies is the space density of the 
 galaxies in that sample as a function of absolute luminosity. 
 In theory, if one could select a 
  complete sample of  galaxies over a sufficiently large 
 section of the  local Universe such that large-scale structure and 
 clustering effects could be ignored then the  LF which 
 could be constructed would provide a definitive measure of the 
 local luminosity density. In practice the many studies of 
 the field population LF (Loveday et al. 1992; 
 Lin et al. 1997; Marzke et al. 1997; Bromley et al. 1998; Muriel, Valotto \& Lambas 1998; 
 Zucca et al. 1997; Folkes et al. 1999) 
  have produced significantly differing 
 results and have generally not been complete to very faint limits. 
  An alternative approach is to seek  for clues about
 galaxy formation and evolution by 
   constructing the LF for 
 particular environments and then studying the difference in the derived LFs between
 environments. 
 The many recent studies of the  LFs of particular clusters (e.g. Driver et al. 1994; 
 de Propris et al. 1995; Lobo et al. 1997; Wilson et al. 1997; 
 Valotto et al. 1997; Smith, Driver \& Phillipps 1997; Trentham 1997a, 
 1997b, 1998; Driver, Couch \& Phillipps 1998b; Garilli, Maccagni \& Andreon 1999) 
  fall into this 
 category.

\begin{figure*}
  \epsfig{file=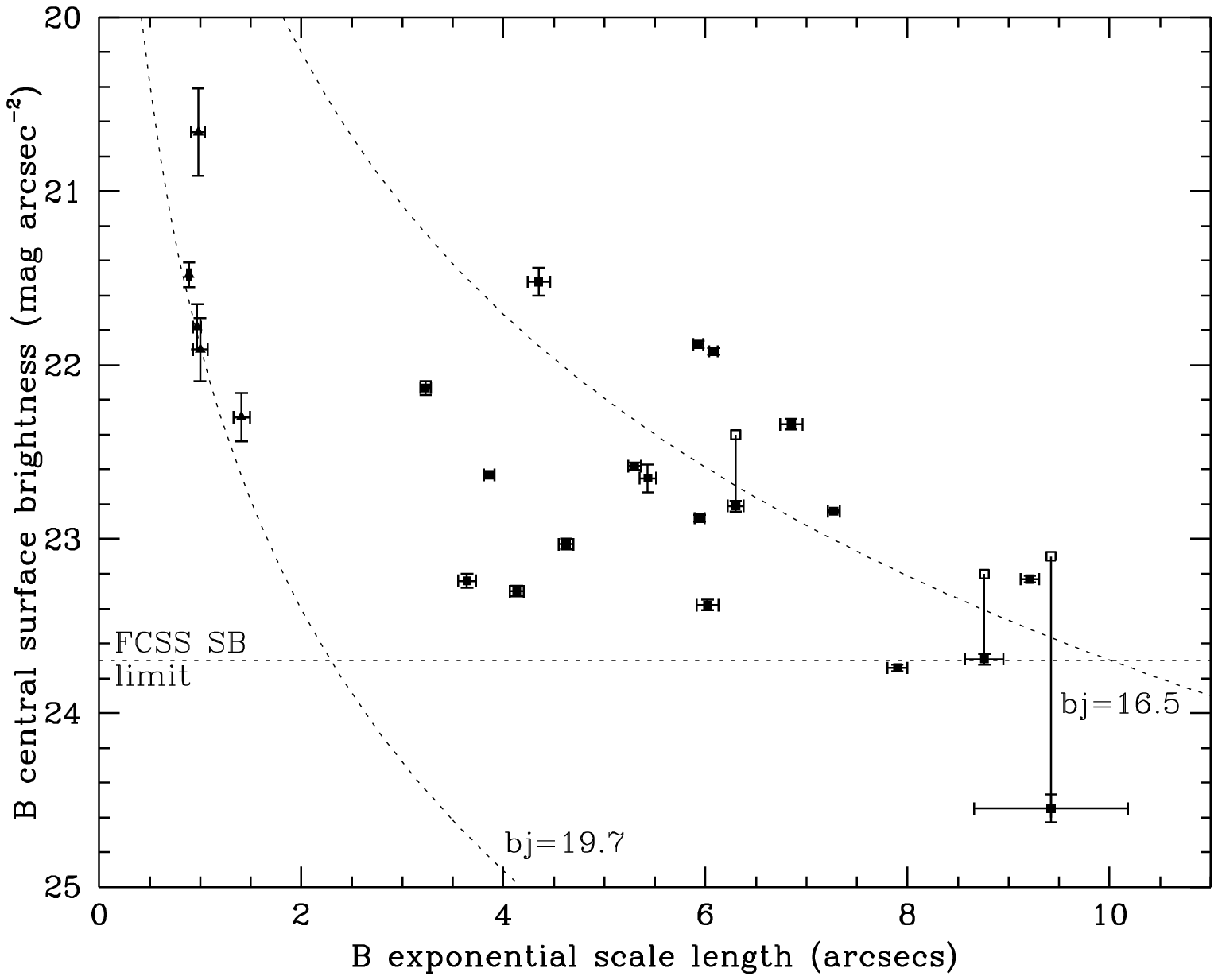}
  \caption{Plot of  B-band extrapolated central surface brightness
 against B-band scale-length 
 for the Fornax Cluster sample. 
  The Ultra Compact Galaxies (UCDs) are denoted by filled triangles, the other Fornax cluster galaxies by 
filled squares. The open squares show the observed central surface brightness 
 for those  3 nucleated galaxies where this is significantly brighter than the extrapolated central surface brightness.}
\end{figure*}

  However, it is now well established that 
 the inclusion of a galaxy in an optically 
 selected sample is critically dependent not only on its absolute  
  luminosity but also on its surface brightness (Disney 1976; 
 Phillipps, Davies \& Disney 1990; Impey \& Bothun 1997).  It is also 
 well established that LSBGs exist with a wide range of total 
luminosities. Such galaxies will have been excluded from most determinations 
 of the LF. A more thorough 
 and physically meaningful way to quantify the extragalactic population 
 is the construction of a Bivariate Brightness Distribution (BBD), i.e. 
 the space density of galaxies as a function of their total luminosity and 
 their surface brightness or, equivalently, 
 as a function of their size and surface brightness
 (Boyce \& Phillipps 1995; Dalcanton 1998; Cross \& Driver 2002).
 A consideration of any possible surface brightness--luminosity relation 
  is then  encompassed as part of the wider determination of the BBD. 
  Past attempts at deriving 
  a BBD (e.g. Choloniewski 1985; van der Kruit 1987; 
 Sodre \& Lahav 1993) have generally suffered from poor statistics. 
  The advent of  wide-field imaging surveys and dedicated spectroscopic surveys 
 has made the measurement of the BBD for statistically significant populations 
 possible for the first time (Driver 1999; Cross et al. 2001; 
 de Jong \& Lacey 2000; Blanton et al. 2001).  These authors have all 
 studied the `field' population to relatively bright limits.

A thorough attempt, therefore, to catalogue and quantify the contents of a cluster, 
   requires, as a first stage, the detection of a complete sample of 
 cluster objects to well defined (preferable very faint) limits in absolute luminosity and 
 surface brightness. Secondly, the morphologies of the various objects found need 
 to be determined. Finally, the BBD for each  morphological type  
   can be constructed. Even this could be further refined since, for example, 
  one might expect the morphological composition of the cluster core to differ from its outer 
 regions. Hence, one may wish to study the BBD as a function of cluster radius. 
 
 Whilst we do not yet have a large enough sample of Fornax Cluster objects to undertake a
 full determination of the cluster BBD, we do have a complete sample close to the cluster centre 
 with well defined luminosity and surface brightness limits and can, therefore, 
 draw some general conclusions about the luminosity--surface brightness  or 
 size--surface brightness
 distribution for the first time.

 Fig.~3 shows the size--surface brightness distribution of the Fornax 
 Cluster sample using the B-band data from Table~2 (i.e. plotting 
   $\mu_{0}$ against $\alpha$).
 The limits imposed by the APM magnitude selection limits 
 ($16.5 \leq b_{j} \leq 19.7$) are noted. 
 These boundaries become slightly
fuzzy in terms
of the final photometry since the target selection was in terms of  
 APM $b_j$ magnitudes.
 The actual APM catalogue limit (on isophotal size at a threshold surface 
 brightness)
hardly impinges on our sample, as noted earlier,
since the low surface brightness galaxies which we are biased against
 including because of their small isophotal sizes 
  (see e.g. Davies et al. 1988) have too low a surface brightness
to be successfully observed by 2dF anyway.
 Also noted in Fig.~3 is the position of 
 our absolute  surface brightness limit at 23.7~B~mag~arcsec$^{-2}$. 
  The surface brightness effective limit
is a fuzzy boundary between 23.2 and 23.7~B~mag~arcsec$^{-2}$ where the 2dF 
success rate falls from $\simeq$80 per cent to zero.

Since we are dealing with fibre spectra, it is only the central
region of the object which is observed. This can have a positive effect
on the inclusion of low surface brightness objects in 
 the case where they are nucleated (i.e. dE,Ns)
since the nucleus may be bright enough to obtain a successful spectrum
even if it sits in a very low surface brightness galaxy. 
For example, we noted 
 in Section 3.2 that J034100.8-354433 has a prominent nuclear bulge 
 and that the values of $\mu{o}$ and $\alpha$ included in Table~2 refer only to 
 the outer disk component of this galaxy. In fact, 
 according to the FCC, 12 of our sample are nucleated (see Table 1). However, in 
   our imaging  
   data only 3 (J034023.5-351636, J034100.8-354433 and J034217.3-353226) 
 show an observed central surface brightness significantly 
 brighter than the extrapolated value listed in Table 2. For these 3 
 objects, we show on Fig.~3 the observed as well as the extrapolated central surface 
 brightness. This makes it clear how  J034100.8-354433 in particular 
  has been included in the sample
 despite having an extrapolated central surface brightness of 24.6~B~mags~arcsec$^{-2}$.

 From Fig.~3 it can be clearly seen that the Fornax sample 
 falls into two distinct areas of the size--surface brightness 
 plane. 
Firstly, 
  the UCDs can be seen around $\alpha$$\simeq$1~arcsec ($\simeq$100~pc at the 
distance of Fornax)  between 
 $\mu_{o}$=20.5--22.3~B~mag~arcsec$^{-2}$.
Secondly, the rest of the (non-compact) dwarfs have  $\alpha>$3~arcsec. 
  Above $\alpha$=3~arcsec ($\simeq$300~pc), the cluster sample essentially fills all of the
available ($\alpha$,$\mu_{o}$) parameter space and therefore 
 shows no correlation between the parameters. This non-correlation 
  was claimed previously
on purely photometric grounds by Davies et al. (1988) and Irwin et al. (1990).
However, the  area of parameter space between $\alpha$=1--3~arcsec 
 contains no galaxies. There appears to be a minimum scale size which 
a Fornax dE can have. 

Whilst Fig.~3 re-emphasises that the UCDs appear to form a distinct population
 from the larger `normal' dwarfs, one cannot  extrapolate 
  too much about the relative properties of the UCDs and the 
 non-compact objects from 
  this figure alone.  
   As noted above, due 
  to the unresolved nature of the UCDs in the photographic data, 
  the plotted values  of 
 $\mu_{o}$ and $\alpha$ are merely limits: the UCDs may be smaller and have 
  higher surface brightness than we measure.   
   In fact a preliminary analysis of HST imaging of these objects 
 (Drinkwater et al, in prep.) suggests that the UCDs actually have scale
 sizes as small as 0.1~arcsec and central surface brightnesses up to 3~mag 
 brighter than measured from our photographic data. The bright central surface 
  brightnesses of the UCDs will make them detectable to fainter magnitudes than 
 the non-compact objects and hence the relative number densities of the two populations 
 are not clearly illustrated by Fig. 3. The nature of the UCDs 
  is currently a subject of some discussion  (Drinkwater et al. 2000b; Hilker 
 et al. 1999;  Mieske, Hilker \& Infante 2002), the debate  focussing 
 on whether the UCDs form  a distinct population of galaxies not linked to globular clusters 
   or whether there is a smooth
 transition between both populations. We defer further consideration of this issue  
 until  presentation of our HST results.

 From Fig.~3, we can say little about the area with surface brightness fainter than 23.7~B$\mu$. 
 However, it is noteworthy that we have detected 6 objects with 
 surface brightness between 23.2--23.7~B~mag~arcsec$^{-2}$, despite the completeness of the 
 sample as whole falling from 80 to 0 per cent in this  
  range. 
 We might reasonably expect further dwarfs to lie within this region.

\begin{figure*}
   \epsfig{file=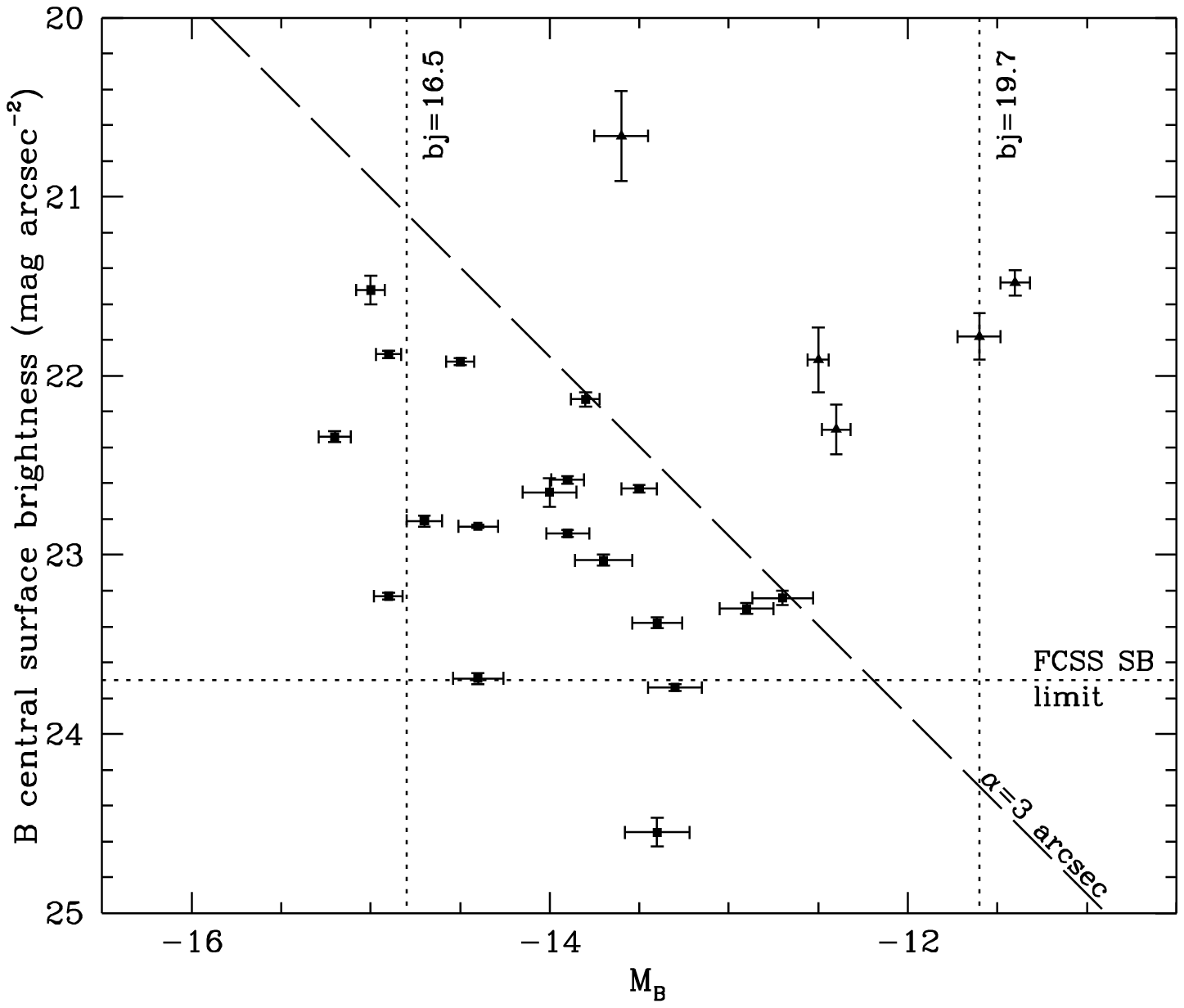}
  \caption{Plot of  B-band  
   extrapolated central surface brightness against 
  B-band absolute magnitude  for the Fornax Cluster sample. 
 The Ultra Compact Galaxies (UCDs) are denoted by filled triangles, the other Fornax cluster galaxies by 
filled squares.}
\end{figure*}

Fig.~4 shows the relationship in its alternative form, showing  
 $B$ central surface brightness against 
 absolute $B$ magnitude. The upper and lower 
  $b_{j}$ magnitude selection limits are once again shown along with 
 the absolute surface brightness  limit at 23.7~B~mag~arcsec$^{-2}$.  
  The outlier FCSS~J034100.8-354433 
  is again seen apparently defying these limits at 
   $\mu_{o}$$\simeq$24.5~B~mag~arcsec$^{-2}$, 
  $M_{B}$$\simeq$--13.4.

Also shown in Fig.~4 is  the locus 
   for a galaxy with a pure exponential profile with a 
 constant scale size of 3~arcsec ($\simeq$300~pc at the distance of 
 Fornax),  assuming 
 a circular image. There is a large area of 
 parameter space to the upper right of this line within which the FCSS 
 would have detected galaxies, 
  but the only galaxies 
 within it are  the UCDs. However, as noted above,  the UCDs may have 
 central surface brightness up to 3~mag brighter than derived from 
 our photometry. If so, then the plotted area of parameter space to the upper right of the 
 $\alpha$=3~arcsec locus would  be completely empty. 
 Below the $\alpha$=3~arcsec locus the sample  fills the entirety of 
  available parameter space. 
 It is notable that this limit in scale-size imposes on the sample 
   the upper-bound of an apparent  surface brightness--magnitude `correlation'. 
   However,  to our limits of $M_{B}$ and $\mu_{o}$ 
  there is no evidence for a lower bound to any such correlation. 
 If such a surface brightness--magnitude correlation does exist 
 for the Fornax dwarfs  it must be fairly broad. For example, 
 at $M_{B}$=--15.0 the surface brightness  distribution is $\ga$2.5~mag in width.

\begin{figure*}
  \epsfig{file=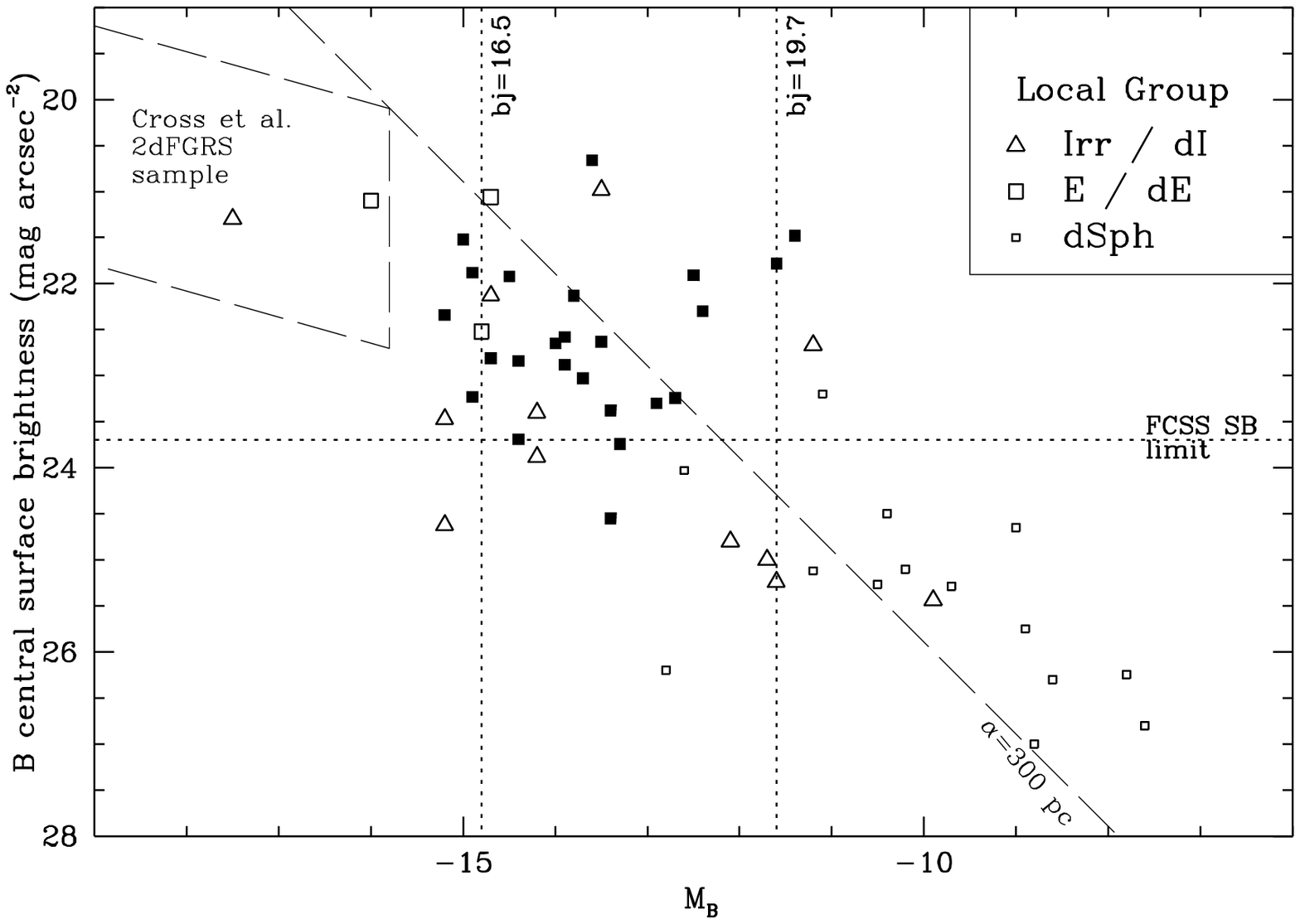}
  \caption{Plot of B-band  
   extrapolated central surface brightness against  B-band absolute magnitude 
   for the Fornax Cluster sample (filled squares), for the 
 Local Group (open symbols -- see key) and for the faint end of Cross et al.'s (2001) 2dFGRS sample.}
\end{figure*}

In Fig.~5 we  again present the Fornax dwarf sample on a plot of
  central surface brightness against absolute luminosity. We also show on this figure the 
 positions of Local Group galaxies taken from Mateo (1998). One needs to be wary 
 of reading too much into this diagram. 
 Whilst we might reasonably 
 expect that Local Group galaxies form the most complete sample of galaxies  at 
 faint luminosities and low surface brightness available,   the Local 
 Group data does not form a sample to strictly defined completeness limits. 
  We are also, of course, comparing 
 galaxies from two different environments (i.e. the centre of the Fornax cluster and 
 the much more diffuse Local Group). The presence of several dIs in the Local Group, where 
 none was found in Field 1 of the FCSS, is testimony to this. In common with the 
 Fornax dEs, all of  the Local Group dEs also have $\alpha$$\ga$300~pc. 
 Most of the Local Group dIs are also constrained by this scale size limit, apart 
 from two outliers, both of which lie in the region of ($M_{B}$,$\mu_{o}$) space 
  occupied by the UCDs. However, as noted above, the UCDs 
   have central surface brightnesses up to 3~mag brighter than derived from our photometry, 
 in which case the Local Group dIs do not in reality lie in a similar region 
 of ($M_{B}$,$\mu_{o}$) space.
 The dSphs of the Local Group 
 do appear to  show a correlation between surface brightness and absolute magnitude, although 
 these objects are generally considered to form a distinct population from the 
 brighter dEs. Many of the Local 
 Group dSphs  have scale-sizes less than 300~pc. 
 Our FCSS data are not 
 sensitive enough to detect any similar objects in Fornax.

We also note on Fig.~5 the region at bright absolute luminosities ($M_{B}<$--16.0) 
 within which the galaxies from the faint end
  of Cross et al.'s (2001)  2dFGRS sample lie. This is 
  one of the largest  samples of bright galaxies yet assembled 
  and formed the basis of 
 Cross et al.'s determination of the BBD
 (see also Cross \& Driver 2002). It should be noted, however, that Cross et al.'s
  sample is not differentiated 
 by morphological type (it includes ellipticals, bulge dominated spirals, disc dominated  
 spirals and irregulars).  In order to compare the Cross et al. sample with our data, 
 we have translated their effective surface brightnesses to central surface brightness 
 by assuming a shift of $\simeq$1.8~mag appropriate for an exponential profile. 
 Within the limits of their sample, 
  Cross et al. found a correlation between surface brightness and absolute magnitude
  such that  $\mu_{e}$ scales as 0.42$M_{B}$. The relationship  
 is around 2.5~mags wide in surface brightness at each absolute magnitude. 
 It is clear from the Fornax and Local Group data that this relationship 
 does not extend to fainter magnitudes. As noted, the  lower limit to the
scale size of the dEs (at $\alpha$$\simeq$300~pc)  creates 
an apparent upper-bound to a surface brightness-magnitude `correlation'. 
 Obviously the  $\mu_{e}$ of this upper-bound scales  as $M_{B}$, i.e.    
   much steeper than the Cross et al. correlation. There is, though, no evidence that 
 there is a low surface brightness bound to any such  correlation for dwarfs. 
 The correlation between surface brightness and absolute magnitude for Local Group dSphs 
 is also much steeper than that from Cross et al.'s bright galaxy sample.

\subsection{Morphology as an indicator of cluster membership}

A prime aim of the FCSS is to test the reliability of 
 using optical morphology as a basis for determining cluster membership. 
 Firstly,  it is worth noting that  only $\simeq$50 per cent of our complete cluster
 sample galaxies actually have $\mu_{0}$$>$22.5~B~mag~arcsec$^{-2}$, the criterion often used
to define a low surface brightness galaxy. In other words, though all
our objects are dwarfs, many of them extreme dwarfs, in terms of luminosity,
half of them are {\em not} low surface brightness galaxies. So, any selection 
 based on the assumption that all cluster dwarfs are going to be  LSBGs  
  will be flawed.  Of course, many
other  LSB dwarfs which are really present will be too faint for us to obtain
a spectroscopic redshift, so the true fraction of non-LSBG dwarfs will
be considerably smaller than 50 per cent, but the important point as regards
morphological (pre)selection is that such dwarfs do exist. 
In addition we find 15 background objects with  $\mu_{0}$$>$22.5~B~mag~arcsec$^{-2}$
 which would morphologically be likely to be assigned to the cluster.

  Our best point of comparison is with 
 Ferguson's (1989) FCC. 
 Ferguson catalogued 2678 objects from photographic plates of an area 
 40~deg$^{2}$ centered on the Fornax cluster. He gave each object 
 one of five classifications: `class 1 - definite cluster member';
 `class 2 - probable cluster member'; `class 3 - possible 
cluster member'; `class 4 - likely background object'; 
 and `class 5 - definite background object'.

Of our complete 
 sample of 24 Fornax Cluster dwarfs,  a total of 17 (71 per cent of our cluster 
 sample) were 
 ascribed by Ferguson to the cluster: 12 as class 1;
    4 as class 2; and 1 as class 3.    
  One of our cluster sample (FCSS~J034159.4-352053) was described as 
  class 5 (`definite background object'). 
 This object has the smallest scale size (3.23~arcsec) of any of the 
 non-UCDs in our sample, although its surface brightness and absolute magnitude 
 are not remarkable compared to the rest of the cluster sample. 
 Another of our sample (FCSS~J033816.7-353027) was listed by 
 Ferguson as a `class 4'  (`possible background object'). 
  This does not have a remarkable surface brightness or absolute magnitude although 
 it has one of the smaller scale lengths of the non-UCD objects 
 (5.43~arcsec). Ferguson did not include any of the UCDs within the FCC either as background galaxies 
 or cluster members.  Presumably these were so compact as to be considered  
 foreground stars.

In addition, within the FCSS survey of Field 1 we found that 
  a further 5 objects which Ferguson had ascribed to the cluster  (1 as class 1;   
   4 as class 2)  are actually background objects.  So within our field and 
 our selection limits, Ferguson ascribed 22 objects to the cluster. FCSS has found that 5 
  of these objects  (i.e. 23 per cent of Ferguson's cluster objects) are actually  background objects. 
   A further 9 objects which were listed by Ferguson as class 3 (`possible cluster 
 member') but included in his table of background galaxies, are confirmed 
 by FCSS as background objects. 
 
 To summarise:  Ferguson did not include any of the FCSS's  UCDs within 
 the FCC (either as 'cluster' or 'background' objects), 
 presumably assuming them to be stars. Ferguson did include within the FCC  all 19 of the 
     non-compact  cluster dwarfs from the FCSS: however, he mis-classified 2 of these as `background' objects. 
  Hence, the FCC contains 89 per cent of the  non-compact
   dwarfs  within the FCSS cluster sample (but only 
 71 per cent if one considers all dwarfs in the FCSS).  However, the FCC contains a further 
 5 objects as `cluster' members which have been  shown by the FCSS to be background objects: 
 i.e.   within the FCSS's field  
   and selection limits, 23 per cent of those objects 
   described as  cluster members by the FCC 
   are shown by the FCSS to be background objects. 
  This clearly illustrates that attempts to determine cluster membership solely 
 on the basis of observed morphology can produce significant errors: firstly because many compact galaxies will
 be mis-classified as stars; and secondly because many  non-compact background objects will be 
  mis-classified as cluster members and a smaller number of  cluster members will be 
 mis-classified as background objects.

\section{Conclusions}

We have obtained spectra for a morphologically unbiased set of cluster 
dwarf galaxies as part of the  Fornax Cluster Spectroscopic Survey.
The present paper provides the photometric study of these confirmed
cluster dwarfs. We find that, despite their low luminosities, not all
the dwarfs also have low surface brightness. There is therefore a wide
spread of parameters in the space of surface brightness versus luminosity
 or scale size. The only correlation evident in these plots is the one
in the luminosity--surface brightness plane caused by the absence of
`normal' dwarfs with scale sizes below about 300~pc. Even then, a separate
class of Ultra-Compact Dwarfs fills some of this otherwise empty area of parameter
space. Morphological 
 selection of cluster dwarfs solely on the grounds
of their low surface brightness may therefore overlook a significant
number of dwarf galaxies which are virtually impossible to discriminate from
background, larger galaxies on appearance alone.

The data presented within this paper result from the completion of only one of the 
 intended 4 FCSS fields.  We now have full data for the second field and are in the process 
 of reducing these. The completion of all 4 fields will not only improve the statistics of 
 our results  but will also enable us to study  the distribution 
 of dwarf galaxies  
 of different morphologies (i.e. dE, dI, UCD) 
  as a function of position within the cluster. Such studies will 
 be assisted by the 
 deep  CCD g,r,i,z  imaging of the cluster region which we have now obtained
 using the 4.0-m Blanco Telescope  at the 
 Cerro Tololo Inter-American Observatory. These data are currently  being reduced.

Important  scientific advances  could also potentially be made by improving the 
 sensitivity of our spectroscopy to fainter luminosities and/or surface brightnesses. 
 This can be clearly seen by studying Figs.~4 and 5. 
  Using the  CCD data we should be able to define a sample to a considerably fainter 
   $B$ magnitude limit.  The standard 2dF set-up would be adequate to obtain spectra of 
  the higher surface brightness objects in such a sample.  This would include many potential 
 UCDs. However, even bigger advances will require us to obtain spectra for objects 
  at fainter surface brightness limits than the present limit of 23.7~B~mag~arcsec$^{-2}$.  
   This possibility now exists with the commissioning of multi-object spectrographs on 
 the new generation of 8-m telescopes.

 The number of cluster objects found in Field 1 is dwarfed by the number of confirmed background 
 objects (1175). Such a sample is potentially of great value to studies of the BBD of field 
 galaxies. Although the present magnitude limit is only slightly lower than that of the 2dFGRS 
  used by Cross  et al. (2001) for their determination of the BBD,  the FCSS background sample  
 has the advantage that  the FCSS sources were not pre-selected as galaxies. 
 Hence, our sample will  contain a fairer representation of compact objects in the background 
 than Cross et al.'s (see Paper~II). For the lower surface brightness objects 
 in our sample we  have also used longer exposure times than the 2dFGRS, so FCSS extends to fainter 
 surface brightness than 2dFGRS.
 However, if the FCSS sample could be pushed to even fainter luminosity 
 and surface brightness limits then, not only will this enable us to study the cluster BBD better, but 
  the resulting sample will be ideal for studying the field BBD to fainter surface brightness 
  limits than  yet obtained.  For example, Cross et al.'s correlation between luminosity 
 and surface brightness has only been tested between  --24$<M_{B}<$--16.0 and 18.0$<\mu_{e}<$24.5.  
 At present the FSCC is  $\simeq$80 per cent complete to a surface brightness about 0.5~mag fainter than this. 
  Further improvements in the sensitivity of the spectroscopy will enable us to sample whole 
 new regions of ($M_{B}$,$\mu_{o}$)  space.

\section*{Acknowledgments}

This project would not have been possible without the superb 2dF facility
provided by the AAO and the generous allocations of observing time from
PATT and ATAC. We wish to thank the other members of the FCSS Team: 
Jonathon Davies, Elaine Sadler and 
Quentin Parker.   JBJ, PJB and JHD acknowledge support
from the UK Particle Physics and Astronomy Research Council, and MJD 
that from the Australian Research Council.  Part of this work was
done at the Institute of Geophysics and Planetary Physics, under the
auspices of the U.S. Department of Energy by Lawrence Livermore
National Laboratory under contract No.~W-7405-Eng-48.
  This research has made use of the NASA/IPAC Extragalactic Database (NED) which is
operated by the Jet Propulsion Laboratory, Caltech, under agreement with the 
 National Aeronautics and Space Administration.

\label{lastpage}

\end{document}